\documentclass[11pt]{article}

\def\be{\begin{equation}}
\def\ee{\end{equation}}
\def\ba{\begin{eqnarray}}
\def\ea{\end{eqnarray}}
\def\nn{\nonumber}
\def\lb{\label}
\def\dfrac{\displaystyle\frac}
\def\bb{\bibitem}

\def\bA{{\cal A}}
\def\H_n{h}
\def\S_n{s}

\begin{document}

\begin{titlepage}
\date{22 June 2023}
\title{\begin{flushright}\begin{small}    LAPTH-022/23
\end{small} \end{flushright} \vspace{1.5cm}
The first law for stationary axisymmetric \\ multi-black hole systems}
\author{G\'erard Cl\'ement$^a$\thanks{Email: gclement@lapth.cnrs.fr} and
Dmitri Gal'tsov$^{b,c}$\thanks{Email: galtsov@phys.msu.ru} \\ \\
$^a$ {\small LAPTh, Universit\'e Savoie Mont Blanc, CNRS, F-74940  Annecy, France} \\*
$^b$ {\small Department of Theoretical Physics, Faculty of Physics,}\\
{\small Moscow State University, 119899, Moscow, Russia }\\
$^c$ {\small  Kazan Federal University, 420008 Kazan, Russia}}
\maketitle

\begin{abstract}
In the framework of Einstein-Maxwell theory, we consider collinear arrays of rotating
electrically charged black holes connected by Misner-Dirac strings carrying gravimagnetic
and magnetic fluxes. The first law of mechanics for
these systems is derived. It involves as dynamical variables the areas, angular momenta
and electric charges of the various Killing horizons --- black holes and Misner strings.
When the gravimagnetic fluxes all vanish, the first law reduces to a form where the
dynamical variables associated with the strings are the string tensions and magnetic fluxes.
This form is not generically invariant under electric-magnetic duality.
\end{abstract}
\end{titlepage}
\setcounter{page}{2}

\setcounter{equation}{0}

\section{Introduction}

Recently, extending previous work by Tomimatsu \cite{tom83,tom84}, we proposed a new approach \cite{Clement:2017otx, smarrnut} to derive mass formulas for stationary axisymmetric solutions of the Einstein-Maxwell equations describing systems of dyonic black holes endowed with electric and magnetic charges, mass, and gravimagnetic (or NUT) mass. The total Komar mass and angular momentum and the total electric charge, given by surface integrals at spatial infinity, are reexpressed as the sum of surface integrals over all the Killing horizons of the multi-black hole configuration. These include both timelike Killing horizons --- black hole event horizons, and spacelike Killing horizons --- Misner strings extending from one black hole to another or to infinity. For each Killing horizon one thus defines local dynamical variables --- mass, angular momentum, electric charge, and surface area, which combine to yield a Smarr relation of the Kerr-Newman type. Adding together these various local Smarr relations, one obtains a global Smarr relation expressing the total mass of the system in terms of all the other local dynamical variables.

In this approach, neither the Dirac quantization of the magnetic charge nor the Misner time periodicity \cite{Misner:1963fr} conditions are imposed, so the Dirac string {\em is} a vector field singularity, and the Misner string {\em is} a spacetime singularity. They have distributional energy-momentum tensors that contribute to the general mass formulas and the first law of black hole mechanics. Together with struts in multicenter solutions, they constitute a set of defects on the polar axis, treated in the same way as black hole horizons. Note that in the case of multiple black holes with different Hawking temperatures, thermodynamic language becomes ambiguous, so we prefer to talk about black hole mechanics, closer in spirit to the original Smarr paper \cite{Smarr:1972kt}  and  the preceeding work by Christodoulou \cite{ Christodoulou:1970wf} and Bekenstein \cite{Bekenstein:1972tm} (see also \cite{Bardeen:1973gs}) before Hawking's entropic interpretation was introduced \cite{Hawking:1976de} and extended to Misner strings  \cite{Hunter:1998qe,Hawking:1998jf}.
This approach has full generality and includes many particular cases which were considered previously in isolation.

Let us emphasize that the Smarr relations thus derived from first principles differ markedly from those proposed elsewhere. The Einstein-Maxwell Lagrangian and its supergravity generalizations, including scalar fields, are locally invariant under electromagnetic S-symmetry, so it is commonly expected that Smarr relations should involve symmetrically black hole electric and magnetic charges. In previous literature, the existence of S-invariant Smarr formulas for dyonic black holes has been discussed in various ways (see recent papers \cite{Mitsios:2021zrn,Ortin:2022uxa,Ballesteros:2023iqb} and references therein). However the nature of electric and magnetic charges is different: the former are associated with conserved currents, while the latter are topological, and give rise to Dirac strings which break the local S-symmetry. So the dyon solutions with Dirac string are not globally S-invariant. This is reflected in the Smarr relations, which involve only electric charges, both for black holes and Dirac strings (which in the axisymmetric setting coincide with Misner strings), but not magnetic charges.

The situation is even worse for the gravimagnetic charges. In the case of NUTty black holes, there have been a number of attempts (see e.g. \cite{bordo2020} and refs. therein) to write down mass relations involving a black hole Misner charge and a conjugate Misner potential. Our Smarr relations do not explicitly involve black hole gravimagnetic or NUT charges. These give hidden contributions to the Misner string masses and angular momenta, which depend on the gravimagnetic fluxes --- differences between the magnetic charges of the two black holes connected by the Misner string.

The purpose of the present paper is twofold. First, the existence of a global mass relation of the Smarr type for interacting systems of black holes and Misner strings suggests that these systems may satisfy a more fundamental differential first law of dynamics. Indeed, for specific NUTty configurations with one or two black holes, a differential mass formula involving the dynamical variables associated with all the Killing horizons has been shown to be valid \cite{smarrnut,bordo2020,dinut}. We shall derive here the general form of the differential first law valid for all asymptotically flat stationary interacting systems of black holes and Misner strings.

Another issue is to understand what happens to the global Smarr relation and differential first law when the NUT charges go to zero so that Misner strings disappear and the axis condition is satisfied. Generically, stationary multi-black hole configurations require the existence of
interconnecting struts or cosmic strings whose tension balances the inter-black hole forces. It has been shown in special cases \cite{herdeiro2010,krtous,gregory} that these systems obey a first law of dynamics where the string variables are the string tension and the conjugate string thermodynamical length. We shall derive here the general form of the Smarr relations and first law valid for interacting systems of rotating dyonic black holes. In the dyonic case, the black hole electric charges enter the black hole Smarr relations, while the black hole magnetic charges enter the Dirac string Smarr relations through the magnetic flux from one black hole to the next. Again, S-symmetry is generically broken.

In the next section, we summarize the Komar-Tomimatsu approach to the Smarr formula,
as generalized in \cite{smarrnut} to systems of black holes and Misner strings.
The first law for such systems is derived in the third section, where
several limiting cases are also discussed. We show in section 4 that when the gravimagnetic fluxes
go to zero, the reduced form of the first law involves,
along with the usual black hole dynamical variables, the string tensions and the magnetic fluxes
carried by the connecting strings. Applications  to the dyonic IWP dipole, to a particular interacting dyonic two-black hole system, and to
the Bonnor magnetostatic solution are presented in the fifth section. We close with a brief discussion.

\setcounter{equation}{0}
\section{The Smarr formulas for multi-black hole systems with NUTs}
We first review briefly the derivation of the Smarr formulas for collinear systems of black holes
and Misner strings, as given in \cite{smarrnut}. The metric for stationary axisymmetric systems can be
parameterized in Weyl coordinates $\rho, z$ as
\begin{equation}\label{Gzr}
 ds^2 = G_{ab}(\rho, z) dx^a dx^b + e^{2\nu(\rho,z)}(d\rho^2 +  dz^2),
\end{equation}
where $a,b=0,1$, with e.g. $x^0 = t, x^1 = \varphi$, and $ \rho = \sqrt{|\det G|}$. The Gram matrix $G(\rho,z)$ is
degenerate on the axis $\rho = 0$, which can be divided into rods \cite{Harmark:2004rm}
where it is singly degenerate, separated by turning points $z_n$ where it is doubly degenerate.
On each rod $z_n < z < z_{n+1}$, $G$ has a constant null eigenvector belonging to the Killing vector two-space (rod direction),
$l_n = \partial_t + \Omega_n\partial_\varphi$,
\begin{equation}
 G_{ab}(0, z)l_{n}^b = 0,
\end{equation}
where $\Omega_n$ is the constant angular velocity of the rod. In Lorentzian spacetime the norm $l_n^2$ of this vector
can be negative, positive or zero for $\rho > 0$, the associated rod being qualified as timelike, spacelike or null.
We shall consider in this paper an asymptotically flat configuration (no strings extending to infinity) of
$2N-1$ rods ($N\ge1$ integer),
$N$ timelike rods (black hole horizons) $H_n =[z_{2n-1},z_{2n}]$ ($n=1 \cdots N$), and  $N-1$ spacelike rods
(Misner strings) $S_n =[z_{2n},z_{2n+1}]$ ($n=1 \cdots N-1$). The  spacetime Killing vectors associated with the rods share with them the same causal nature (timelike for horizons and spacelike for defects) and they become null on the rods themselves.
Thus both timelike and spacelike rods are  Killing horizons with certain surface gravities, which together with the angular velocities $\Omega_n$ are constant along the rods.

The total Komar mass, angular momentum and electric charge of a stationary axisymmetric configuration
are given by the integrals over $\Sigma_\infty$:
 \be
M = \frac1{4\pi}\oint_{\Sigma_\infty}D^\nu k^{\mu}d\Sigma_{\mu\nu}, \quad
J = -\frac1{8\pi}\oint_{\Sigma_\infty}D^\nu m^{\mu}d\Sigma_{\mu\nu},\quad
Q=   \frac1{4\pi}\oint_{\Sigma_\infty} F^{\mu\nu} d\Sigma_{\mu\nu},
\lb{koMJ}
 \ee
where $k^\mu = \delta^\mu_t$ and $m^\mu = \delta^\mu_\varphi$ are
the Killing vectors associated with time translations and rotations
around the $z$-axis, $D^\nu$ is the covariant derivative and $F^{\mu\nu}$ is the Maxwell tensor.
Using the Einstein-Maxwell equations and the Gauss-Ostrogradsky theorem, the Komar charges can be
expressed as the sums of integrals over the various rods $\Sigma_n$ (two-surfaces spanned by the
coordinates $z, \varphi$):
 \be
M = \sum_n M_n, \quad J = \sum_n J_n, \quad Q  = \sum_n Q_n,
 \ee
with
\ba
M_n &=& \frac1{8\pi}\oint_{\Sigma_n}\left[g^{ij}g^{ta}\partial_j g_{ta}
+2(A_t F^{it}-A_\varphi F^{i\varphi})\right]d\Sigma_i, \nn\\
J_n &=& -\frac1{16\pi}\oint_{\Sigma_n}\left[g^{ij}g^{ta}
\partial_jg_{\varphi a} +4A_\varphi F^{it}\right]d\Sigma_i, \nn\\
Q_n &=& \frac1{4\pi}\oint_{\Sigma_n}F^{ti}d\Sigma_i .\lb{rodcharge}
\ea

Tomimatsu showed in \cite{tom83,tom84} how to simplify the computation of these rod Komar charges
by using the fact that the rods are Killing horizons. Actually Tomimatsu considered only black hole
event horizons, but we showed in \cite{smarrnut} that the same formulas were valid for all Killing horizons
-- black hole horizons and Misner strings. Introducing for the metric and electromagnetic one-form the
axisymmetric Weyl-Papapetrou parametrization
 \ba\lb{weyl}
ds^2 &=& -F(dt-\omega d\varphi)^2 + F^{-1}[e^{2k}(d\rho^2+dz^2)+\rho^2d\varphi^2], \nn\\
A &=& vdt + A_\varphi d\varphi,
 \ea
the imaginary parts of the complex electromagnetic and gravitational Ernst potentials,
$u = {\rm Im}\,\psi$ (the scalar magnetic potential) and $\chi = {\rm Im}\,{\cal E}$
(the twist potential) are defined up to an additive constant by the dualizations
 \ba\lb{dual}
\partial_i u &=& F \rho^{-1}\epsilon_{ij}\left(\partial_j A_\varphi + \omega  \partial_j v  \right), \\
\partial_i\chi &=& -F^2\rho^{-1}\epsilon_{ij}\partial_j\omega + 2(u\partial_i v - v\partial_i u)\,,
\ea
$i=1,2$, with $x^1=\rho$, $x^2=z$. Using these to evaluate the integrands in (\ref{rodcharge}) in the limit
$\rho\to0$ leads (after correcting in \cite{Clement:2017otx} the Tomimatsu formula for the horizon mass) to
 \ba
M_n &=& \frac1{8\pi}\int_{\Sigma_n}\left[\omega\partial_z\chi +
2\partial_z(A_\varphi\,u)\right]dzd\varphi, \nn\\
J_n &=& \frac1{16\pi}\int_{\Sigma_n}\omega\left[-2 + \omega\partial_z\chi + 2\partial_z(A_\varphi\,u) - 2\omega
\Phi\partial_z u\right]dzd\varphi, \nn\\
Q_n &=& \frac1{4\pi}\int_{\Sigma_n} \omega\partial_z u\,dzd\varphi. \lb{tomsurf}
 \ea

Taking into account the constancy over each Killing horizon of the horizon angular velocity
$\Omega_n = \omega_n^{-1}$ and of the electric potential $\Phi_n$ in the horizon corotating frame
  \be\lb{Phi}
\Phi_n = - [v + \Omega_nA_\varphi]_n ,
 \ee
the surface integrals (\ref{tomsurf}) can be evaluated as
 \ba
M_n &=& \frac14\omega_n\delta_n\chi +\frac12 \delta_n(A_\varphi\,u), \lb{Mi} \\
J_n &=& \frac14\omega_n\left\{ -L_n + \omega_n \left( \delta_n\chi/2-\Phi_n\delta_n u\right)+
 \delta_n(A_\varphi\,u)\right\}, \lb{Ji} \\
Q_n &=& \frac12\omega_n \delta_n u , \lb{Qi}
 \ea
where for any function of two variables $f(\rho,z)$ the quantity $ \delta_n f\equiv f(0,z_{n+1})-f(0,z_{n})$ is  its
variation between two ends $z_{n+1}$ and $z_n$ of the rod $n$, while $L_n\equiv z_{n+1}-z_n$ is the rod's length  (for more details, see \cite{Clement:2017otx}). The rod length is related to the
rescaled surface area of the Killing horizon
\be\lb{Ar1}
\bA_n =\dfrac1{8\pi}\oint d\varphi\int_{z_{n}}^{z_{n+1}} \sqrt{|g_{zz}g_{\varphi\varphi}|} dz =
\dfrac14\int_{z_n}^{z_{n+1}}\vert e^k\omega \vert dz
\ee
by
\be\lb{Li}
 L_n = 4\kappa_n\bA_n,
\ee
where
\be \lb{k}
\kappa_n = \vert  e^{-k_n}\Omega_n\vert
\ee
is the constant Killing horizon surface gravity.

The relations (\ref{Mi}), (\ref{Ji}), (\ref{Qi}) can be linearly combined to yield the Smarr relation
 \be\lb{smarri}
M_n = 2\Omega_nJ_n + 2\kappa_n\bA_n + \Phi_nQ_n
 \ee
valid for each Killing horizon, black hole event horizon or Misner string. Finally, adding together
these individual Smarr relations and splitting summation indices $n$ in two sets $h,\;s$ enumerating black hole and string contributions respectively, we obtain the global Smarr relation for the system:
 \be\lb{gsmarr}
M = \sum_{h=1}^N\left(2\Omega_{h}J_{h} + 2\kappa_{\H_n}\bA_{\H_n} + \Phi_{\H_n}Q_{\H_n}\right) +
\sum_{s=1}^{N-1}\left(2\Omega_{\S_n}J_{\S_n} + 2\kappa_{\S_n}\bA_{\S_n} + \Phi_{\S_n}Q_{\S_n}\right) ,
 \ee
 where the first sum relates to constituent black holes, and the second to defects.

\setcounter{equation}{0}
\section{The first law for NUTty multi-black holes}

The first law for such a system is
 \be\lb{gfirst}
dM = \sum_{h=1}^N\left(\Omega_{\H_n}dJ_{\H_n} + \kappa_{\H_n} d\bA_{\H_n} + \Phi_{\H_n}dQ_{\H_n}\right) +
\sum_{s=1}^{N-1}\left(\Omega_{\S_n}dJ_{\S_n} + \kappa_{\S_n} d\bA_{\S_n} + \Phi_{\S_n}dQ_{\S_n}\right) ,
 \ee
where $M$ is the total mass (sum of the horizon and string masses). It expresses the differential of the total mass (energy) in terms of differentials of extensive (additive) quantities $J_i, \;{\cal A}_i, \;Q_i$.

To derive it from the Smarr law (\ref{gsmarr}), we use the scaling method as applied in \cite{gauntlett}
and \cite{townsend}. The most general solution for a stationary axisymmetric system of $N$ black holes
depends on $3(2N-1)$ parameters: $3$ parameters for each black hole, associated with black hole mass $m_h$,
rotation parameter $a_h$ and electric charge $q_h$; and $3$ parameters for each of the $N-1$ interconnecting strings,
the string length $\delta_{\S_n}z$ (or the distance between two adjacent black holes), and the gravimagnetic and magnetic fluxes
$\delta_{\S_n}\chi$ and $\delta_{\S_n}u$ exchanged between two adjacent black holes. The total mass of the system depends therefore on
$3(2N-1)$ independent variables, which may be chosen to be the angular momentum, rescaled area, and electric charge
of the $(2N-1)$ Killing horizons:
 \be
M = M(J_i,\,\bA_j,\,Q_k) \qquad (i,j,k = 1, \cdots 2N-1) .
 \ee

Under changes of the length scale $L$, the angular momenta and rescaled areas scale as $L^2$, while the
electric charges and the total mass scale as $L$. Thus, by Euler's theorem on weighted homogeneous functions,
 \ba
M &=& \sum_{h=1}^N\left(2J_{\H_n}\dfrac{\partial M}{\partial J_{\H_n}} + 2\bA_{\H_n}\dfrac{\partial M}{\partial \bA_{\H_n}}
+ Q_{\H_n}\dfrac{\partial M}{\partial Q_{\H_n}}\right) \nn\\ &&
+ \sum_{n=1}^{N-1}\left(2J_{\S_n}\dfrac{\partial M}{\partial J_{\S_n}} + 2\bA_{\S_n}\dfrac{\partial M}{\partial \bA_{\S_n}}
+ Q_{\S_n}\dfrac{\partial M}{\partial Q_{\S_n}}\right) .
 \ea
Comparing with (\ref{gsmarr}), we are led to identify
 \be
\dfrac{\partial M}{\partial J_i} = \Omega_i, \quad \dfrac{\partial M}{\partial \bA_i} = \kappa_i, \quad \dfrac{\partial M}{\partial Q_i} = \Phi_i
 \ee
for all Killing horizons, $i=\H_n$ or $\S_n$, leading to the first law (\ref{gfirst}).

An alternative derivation could follow the classical proof of the first law of black hole mechanics \cite{Bardeen:1973gs},
extended in \cite{carter} to the case of electrically charged black holes, and later reformulated,
with weaker assumptions in \cite{compere}. This last compact derivation does not assume any specific invariance properties
or any specific topology of the horizon, but does assume that the Killing horizon is connected. So at the moment it is not clear
how to generalize this to a multi-connected Killing horizon with different values for the angular velocities and surface gravities
of the various connected components.

\subsection{Special cases}

Now we discuss several exceptional limiting cases where, although the Tomimatsu derivation does not go through, the Smarr relations (\ref{smarri})
and the first law (\ref{gfirst}) are still valid.

a) \underline{Non-rotating black holes}. In this case, the Weyl metric function $F(\rho,z)$ vanishes on the
event horizon rods $\H_n$, so that the only surviving element of the Gram matrix $G(\rho,z)$ on the rod is
$G_{11} =$ lim$_{\rho\to0}(F_{\H_n}^{-1}\rho^2)$, leading to
\be
\Omega_{\H_n} = 0.
\ee
The Tomimatsu relations (\ref{Mi})-(\ref{Qi}) can no longer be used to evaluate the event horizon Komar charges, which nevertheless
satisfy the Smarr relations and the first law with $\Omega_{\H_n} = 0$.

b) \underline{Critically rotating strings}. Under the transformation to a rotating reference frame $d\varphi \to d\varphi' = d\varphi + \Omega dt$, the Weyl metric functions $F$ and $\omega$ transform as
 \be
F' = F(1+\Omega\omega)^2, \qquad \omega' = \frac{\omega}
{(1+\Omega\omega)}.
 \ee
For the critical value of the rotation velocity $\Omega = - \Omega_{\S_n}$, $F'$ and $F'\omega'$ vanish on the Misner string rod $S_n$,
so that the only surviving element of the transformed Gram matrix is $G'_{11} = -F_{\S_n}\omega_{\S_n}^2$, leading to
\be
\Omega'_{\S_n} = 0.
\ee

c) \underline{Extreme black holes}. In this case, the event horizon rod reduces to a single point $H_n$. However the
event horizon is still a topological sphere, with a finite surface area. As discussed in \cite{Clement:2017kas,Clement:2018nmt},
the point $H_n$ can be blown up to a finite segment $H'_n$ by transforming from the Weyl coordinates to new appropriate coordinates.
The event horizon Komar charges can be evaluated from the relations (\ref{Mi})-(\ref{Qi}), where $\delta_n f$ is now the
difference between the values of the function $f$ between the two ends of the segment $H'_n$, and $L_n=0$. These satisfy the Smarr relations (\ref{smarri}) and the first law (\ref{gfirst}) with
\be
\kappa_{\H_n} = 0
\ee
from (\ref{Li}).

d) \underline{Contiguous black holes}. In the symmetrical limiting case where the Misner string between two black holes reduces to a single point $S_n$, one must set
\be
\kappa_{\S_n} = 0
\ee
in the mass relations.

The first law (\ref{gfirst}) for systems of NUTty black holes was first checked in \cite{dinut} on the example of
a system of two electrically neutral black holes with equal masses and opposite NUT charges. We shall give in Sect. 5
two other examples.

\setcounter{equation}{0}
\section{The first law for NUTless dyonic multi-black holes}

A last but important special case is that of the NUTless limit $\omega_{\S_n} \to 0$. In this case, the string angular momenta $J_{\S_n}$ and electric charges $Q_{\S_n}$ vanish, as well as the string rescaled areas
\be
\bA_{\S_n} = \frac14 |\omega_{\S_n}|\lambda_{\S_n},
\ee
where we have defined the string thermodynamic length \cite{appels,krtous}
\be\lb{lambdaS}
\lambda_{\S_n} \equiv L_{\S_n}\,e^{k_{\S_n}}.
\ee
On the other hand, the conjugate variables $\Omega_{\S_n}$, $\Phi_{\S_n}$ and $\kappa_{\S_n}$ diverge in this limit,
so that a finite reformulation of the string contributions to the first law should be possible.

From (\ref{Phi}) we see that $A_{\varphi}$ is constant on the string $\S_n$ in the limit $\omega_{\S_n} \to 0$, with
the value
\be\lb{AphiS}
A_{\varphi}\vert_{\S_n} = -\omega_{\S_n}\Phi_{\S_n} \equiv 2P_{\S_n}.
\ee
In the case of a dihole system ($N=2$), this corresponds to the magnetic flux along the string between the black hole $H_1$
of magnetic charge $P_{S_1}$ and the black hole $H_2$ of magnetic charge $-P_{S_1}$. In the general case $N>2$, the magnetic fluxes
along the different strings have no reason to have the same value, so that the notion of black hole magnetic charge becomes ill defined,
and must be traded for that of string magnetic flux.

Inserting (\ref{AphiS}) in (\ref{Ji})and using (\ref{lambdaS}), we find that in the NUTless limit the angular momentum of the string $S_n$
goes to the value
\be
J_{\S_n} = \omega_{\S_n}\left[P_{\S_n}\delta_{\S_n}u - \dfrac14\lambda_{\S_n} e^{-k_{\S_n}} \right].
\ee
Using this together with $Q_{\S_n}=(1/2)\omega_{\S_n}\delta_{\S_n}u$, $\bA_{\S_n}=(1/4)|\omega_{\S_n}|\lambda_{\S_n}$ and $\kappa_{\S_n} =
|\Omega_{\S_n}|e^{-k_{\S_n}}$, we obtain for the contribution of the $n$th string to the first law:
\ba
X_{\S_n} &\equiv& \Omega_{\S_n}dJ_{\S_n} + \kappa_{\S_n} d\bA_{\S_n} + \Phi_{\S_n}dQ_{\S_n} \nn\\
&=& \Omega_{\S_n}\left[d(\omega_{\S_n}P_{\S_n}\delta_{\S_n}u) - \dfrac14d(\omega_{\S_n}\lambda_{\S_n}e^{-k_{\S_n}}) \right. \nn\\
&& \left. - P_{\S_n}d(\omega_{\S_n}\delta_{\S_n}u) + \dfrac14e^{-k_{\S_n}}d(\omega_{\S_n}\lambda_{\S_n}) \right] \nn\\
&=& \Psi_{\S_n}dP_{\S_n} - \lambda_{\S_n}d\mu_{\S_n},
\ea
where
\be
\Psi_{\S_n} \equiv \delta_{\S_n}u
\ee
is the magnetic potential difference between the two ends of the string, and
\be\lb{strut}
\mu_{\S_n} \equiv -\dfrac14(1-e^{-k_{\S_n}})
\ee
is the positive strut tension \cite{krtous}, opposite to the string tension. The resulting first law is
 \be\lb{dyfirst}
dM = \sum_{h=1}^N\left(\Omega_{\H_n}dJ_{\H_n} + \kappa_{\H_n} d\bA_{\H_n} + \Phi_{\H_n}dQ_{\H_n}\right) +
\sum_{s=1}^{N-1}\left(\Psi_{\S_n}dP_{\S_n} - \lambda_{\S_n}d\mu_{\S_n}\right).
 \ee

This form of the first law for NUTless systems was first derived for a system of two black holes with the same temperature in \cite{herdeiro2010}. Using explicit solutions, it was shown to be also valid for a system of two Reissner-Nordstr\"om black holes with different temperatures in \cite{krtous}, and for a collinear system of many Schwarzschild black holes in \cite{gregory}. We have shown here, without relying on explicit solutions, that it applies to a generic collinear system of many rotating dyonic black holes. Its disadvantage, however, is that the second sum is not expressed through differentials of additive quantities.

Likewise, it is straightforward to show that the string Smarr formulas (\ref{smarri}) reduce in the NUTless limit to
\be\lb{smarrs}
M_{\S_n} = \Psi_{\S_n}P_{\S_n}.
\ee
The string tensions (\ref{strut}) are scale invariant and so, by virtue of the Euler theorem, do not contribute to the NUTless Smarr formulas, as previously pointed out in the vacuum case in \cite{krtous}. The value (\ref{smarrs}) for the Dirac string masses generalizes that obtained in \cite{Clement:2017otx} in the case of the dyonic Kerr-Newman black hole.

Contrary to naive expectations, the first law (\ref{dyfirst}) for axisymmetric arrays of NUTless dyonic black holes
is not generically invariant under electric-magnetic duality. It is true that the magnetic flux through the Dirac string
$S_n$ connecting the two black holes $H_n$ and $H_{n+1}$ can be thought of as the difference
\be
2P_{S_n} = P_{H_{n+1}} - P_{H_n}
\ee
between the two black hole magnetic charges, so that the first law (\ref{dyfirst}) can be rewritten in the apparently S-symmetric form
 \be
dM = \sum_{h=1}^N\left(\Omega_{\H_n}dJ_{\H_n} + \kappa_{\H_n} d\bA_{\H_n} + \Phi_{\H_n}dQ_{\H_n} + \tilde\Phi_{\H_n}dP_{\H_n}\right)  -
\sum_{s=1}^{N-1}\lambda_{\S_n}d\mu_{\S_n},
 \ee
with
\be
2\tilde\Phi_{h} = 2u_{h} - u_{h-1} - u_{h+1} + \lambda
\ee
(the gauge parameter $\lambda$ accounting for the constraint $\sum_{h=1}^N P_{\H_n} = 0$). However it is clear that the ``potentials''
$\tilde\Phi_{\H_n}$ thus formally defined do not characterize the black hole horizon $H_n$, as they depend in a non-local fashion on the values of the magnetic potential $u$ on several horizons.

\setcounter{equation}{0}
\section{Examples}

\subsection{The case of the IWP dipole}

A simple but trivial case is that of the IWP linear superposition of two static dyonic extreme
black holes with equal masses and electric charges, and opposite NUT and magnetic charges.
This was discussed by Hartle and Hawking \cite{HH}, and dismissed by them as singular
unless the Misner time periodicity condition is assumed.

The IWP metric is
\be\lb{iwp}
ds^2 = - |U|^{-2}(dt - \vec\omega\cdot d\vec{x})^2 + |U|^{2}d\vec{x}^2,
\ee
where the complex potential $U$ is harmonic,
\be
\nabla^2 U = 0,
\ee
and
\be\lb{dualiwp}
\nabla\wedge\vec\omega = i[\overline{U}\nabla U - U\nabla\overline{U}].
\ee
The corresponding Ernst potentials are
\be
{\cal E} = \frac{2-U}U, \quad \psi = \frac{1-U}U.
\ee

The system under consideration is the dipole superposition of two non-rotating
extreme black holes located at $\rho=0$, $z=\pm\sigma$,
\be
U = 1 + \frac{M_+}{r_+} + \frac{M_-}{r_-},
\ee
with
\be
M_\pm = M \pm i N, \quad r_\pm^2 = \rho^2 + (z\mp\sigma)^2,
 \ee
where $\rho$ and $z$ are the axisymmetric Weyl coordinates, and $r_\pm \ge 0$. Introduce the spheroidal coordinates
 \be
x = \frac1{2\sigma}(r_+ + r_-), \quad y = \frac1{2\sigma}(r_+ - r_-).
 \ee
One can show that $x \ge 1$ and $-1 \le y \le 1$. Then, the potential can be rewritten as
 \be
U =  \frac{m+in}{x+y} + \frac{m-in}{x-y} = \frac{x^2 + 2mx - y^2 - 2iny}{x^2-y^2},
 \ee
with $M = \sigma m$, $N = \sigma n$. Finally, the flat three-metric in (\ref{iwp}) is
 \ba
d\vec{x}^2 &=& d\rho^2 + dz^2 + \rho^2 d\varphi^2 \nn\\
&=& \sigma^2\left[(x^2-y^2)\left( \frac{dx^2}{x^2-1} + \frac{dy^2}{1-y^2}\right) + (x^2-1)(1-y^2)d\varphi^2\right].
 \ea

The duality equations (\ref{dualiwp}) with coordinates ($x, y$) read
\ba
\partial_x\omega &=&  \dfrac{4\sigma n(x^2 + 2mx + y^2)(1-y^2)}{(x^2-y^2)^2}, \nn\\
\partial_y\omega &=&  \dfrac{8\sigma n(x^2-1)(x+m)}{(x^2-y^2)^2}.
\ea
The solution with the boundary condition $\omega \to 0$ for $x \to \infty$ is
 \be\lb{omphi}
\omega = - \frac{4\sigma n(x+m)(1-y^2)}{x^2-y^2}.
 \ee
This vanishes as expected on the two semi-axes $y=\pm1$ ($\rho=0$, $z>\sigma$ and ($\rho=0$, $z<-\sigma$)
extending from the sources to infinity. On the Dirac-Misner string $x=1$ ($\rho=0$, $-\sigma<z<\sigma$),
 \be
\omega_S = - 4\sigma n(1+m),
 \ee
in accordance with (4.24) of \cite{HH}, with $a=2\sigma$.

The string surface gravity and rescaled area are
 \be\lb{kAS}
\kappa_S = |\Omega_S| \equiv |\omega_S|^{-1}, \qquad \bA_S = \dfrac{\sigma}2|\omega_S|.
 \ee
The string electric charge and Komar mass are given in terms of the functions $u_S(y)$ and $A_{\varphi S}(y)$ by
 \be
Q_S = \left.\dfrac{\omega_S}2 u_S\right]_{-1}^1, \quad M_S = \left[\dfrac{\omega_S}4 \chi_S +\dfrac12A_{\varphi S} u_S\right]_{-1}^1,
 \ee
which both vanish, $Q_S = M_S = 0$, because $u_S(y)$ and  $\chi_S(y)$ vanish at both ends of the string. It then follows from
the string Smarr formula that
 \be
J_S = - \dfrac\sigma2 \omega_S = 2\sigma^2 n(1+m).
 \ee

Knowing these and the asymptotic Komar charges  $M=Q=2\sigma m$ and $J = 2\sigma^2 n$, one obtains by difference the horizon Komar charges
 \be\lb{horiwp}
M_{H\pm} = \sigma m, \quad J_{H\pm} = - \sigma^2 mn, \quad Q_{H\pm} = \sigma m.
 \ee
These satisfy the black hole Smarr relation which, because the two constituent black holes at $x=1, y = \pm1$ are non-rotating,
$\Omega_{H\pm} = 0$, and extreme, $\kappa{H\pm} = 0$, reduces to $M_{H\pm} = \Phi_{H\pm} Q_{H\pm}$,
with $\Phi_{H\pm} = - {\rm Re}\,\psi\vert_{H\pm}  = 1$. The horizon masses and electric charges are equal to the bare quantities
$M_\pm$ and $Q_\pm$, while each horizon angular momentum is the well-known angular momentum of an electric charge
$\sigma m$ in the field of a magnetic monopole (the other black hole) of magnetic charge $- \sigma n$.
Note also that the black hole NUT charges
$\pm \omega_S/4$ differ by a linear correction from the ``bare'' NUT charges $\mp \sigma n$.

From these and the proportionality of $J_S$ and $\bA_S$ follow trivially the first laws
 \be
dM_H = dQ_H, \qquad 0 = \Omega_S dJ_S + \kappa_{S} d\bA_{S}
 \ee
satisfied independently by the non-interacting black holes and string.

\subsection{An interacting dyonic system}

Consider now the magnetized rotating solution generated from the static $\gamma=2$
Zipoy-Voorhees vacuum solution in \cite{kerr}. As analyzed in \cite{Clement:2017kas,Clement:2018nmt},
this two-parameter solution describes an interacting system of two co-rotating extreme black holes $H_\pm$
with equal masses and electric charges,
and opposite NUT and magnetic charges, connected by a Dirac-Misner string $S$ which is a critically rotating
cosmic string (such that $\Omega_S = 0$). The Smarr relations (\ref{smarri}) for the horizons and string simplify to
 \ba
M_{H} &=& 2\Omega_{H}J_{H} + \Phi_{H}Q_{H}, \lb{smarrH}\\
M_{S} &=& 2\kappa_S\bA_{S}  + \Phi_{S}Q_{S}, \lb{smarrS}
 \ea
and the first law is
 \be\lb{firsttale}
dM = 2\left(\Omega_{H}dJ_{H} + \Phi_{H}dQ_{H}\right) +
\kappa_S d\bA_{S} + \Phi_{S}dQ_{S}.
 \ee

The form of the solution, depending on a scale parameter $\sigma$ (the half-string length)
and a single dimensionless parameter $q$ (the rotation parameter), or $p>0$ related to $q$ by
\be
q^2+p^2=1,
\ee
is too cumbersome to be given here. We refer the reader to \cite{Clement:2017kas,Clement:2018nmt}.
The horizon and string Komar masses
 \be
M_H = \frac\sigma{p} +\frac{\sigma p}2, \quad M_S = -\sigma p
 \ee
add up to the total mass
 \be
M = \dfrac{2\sigma}{p}.
 \ee
The other horizon and string observables are
 \ba
J_H &=& \frac{\sigma^2}{8qp}\left[2\lambda(p)(2+p^2)+q^2p(1+p)(2-p)\right], \quad
\Omega_H = \frac{q}{\sigma\lambda(p)}, \nn\\
Q_H &=& -\dfrac{\varepsilon\sigma(1+p)}2, \quad
\Phi_H = \dfrac{\varepsilon q^2(2-p)}{2\lambda(p)}, \nn\\
\bA_S &=& \dfrac{\sigma^2 q^3}{4p}, \quad
\kappa_S = \frac{2p}{\sigma q^3}, \nn\\
Q_S &=& \varepsilon\sigma(1+p), \quad
\Phi_S = -\varepsilon,
 \ea
where $\varepsilon=\pm1$, and $\lambda(p)$ is the function
 \be\lb{lambda}
\lambda(p) = \frac{(1+p)(8-4p+5p^2-p^3)}{2p}.
 \ee

With these values it is easy to check (without entering the specific form of $\lambda(p)$) that
the Smarr relations (\ref{smarrH}) and (\ref{smarrS}) hold. The first law (\ref{gfirst}) then
follows trivially if $\sigma$ is varied with $p$ fixed. To ascertain whether it holds when
$p$ is varied for fixed $\sigma$, one must use (\ref{lambda}).
Checking (\ref{gfirst}) then amounts to showing that all the coefficients of a polynomial of sixth order
in $p$ vanish, which is indeed satisfied. Note that
$\partial M_S/\partial p = \Phi_S(\partial Q_S/\partial p)$, so that the first law is not satisfied
separately by the string alone, and so neither by the horizons. This means that the standard first law
of black hole thermodynamics does not hold for NUTty black holes, and its generalization does not
involve the NUT charge, but rather the string observables.

\subsection{The Bonnor dipole}
Our last example concerns the NUTless system described by the Bonnor magnetostatic solution \cite{bonnor66}.
This was shown in \cite{emparan} (see also \cite{davidson}) to represent a system of two black holes with opposite magnetic charges and
degenerate horizons, held apart by a cosmic string. The Bonnor solution, depending on the half-string length
$\sigma$ and the dimensionless parameter $m$ ($0\le m \le 2$), is given in spheroidal coordinates by
 \ba
ds^2 &=& -Fdt^2 + \sigma^2F^{-1}\left[\dfrac{\zeta^4}{(x^2-y^2)^{3}}\left(\frac{dx^2}{x^2-1} +
\frac{dy^2}{1-y^2}\right) + (x^2-1)(1-y^2)d\varphi^2\right], \nn\\
A &=& \dfrac{\sigma m \delta^{1/2}(2x+m)(1-y^2)}{2\zeta} d\varphi, \lb{bon}
 \ea
with
\be
F = \dfrac{\zeta^2}{[\zeta+m(2x+m)/2]^2}, \quad \zeta = x^2-1+\delta(1-y^2),
\ee
where $\delta = 1 - m^2/4$.

The two constituent black holes located at $x=1, y = \pm1$ being non-rotating, extreme, and electrically neutral,
it follows from their Smarr relations (\ref{smarri}) that their Komar mass vanishes,
\be
M_{H_\pm} = 0,
\ee
as found by direct computation from (\ref{Mi}) in \cite{dimagn}. It follows that the mass $M = \sigma m$ of the
Bonnor dipole must be entirely due to the Komar mass of the Dirac string $x=1$. Indeed, knowing the scalar magnetic potential
\cite{dimagn}
\be
u(x,y) = \dfrac{m\delta^{1/2}y}{\zeta+m(2x+m)/2},
\ee
one obtains the magnetic potential difference
\be\lb{psibon}
\psi_S = u(1,1) - u(1,-1) = 2\sqrt{\dfrac{2-m}{2+m}},
\ee
while the magnetic flux is
\be\lb{fluxbon}
P_S = \dfrac12A_\varphi(x=1) = \dfrac{\sigma m}2 \sqrt{\dfrac{2+m}{2-m}},
\ee
leading from (\ref{smarrs}) to
\be
M_S = \sigma m=M.
\ee

The first law (\ref{dyfirst}) reduces in this case to
 \be\lb{firstbon}
dM = \Psi_{S}dP_{S} - \lambda_{S}d\mu_{S}.
 \ee
From the Bonnor metric (\ref{bon}), one obtains for the string $x=1$:
$L_S = 2\sigma$ and $e^{k_S} = \delta^2$, leading to the cosmic string dynamical variables
\be
\lambda_S = 2\sigma\delta^2, \qquad \mu_S =\dfrac14(1-\delta^{-2}).
\ee
It is then a simple matter to check the first law (\ref{firstbon}), which states the dynamical balance
between the cosmic string and the (magnetic) Dirac string. In this context, let us recall that the cosmic
string can be removed by immersing the Bonnor dipole in an external magnetic field \cite{emparan}.

\section{Conclusion}
We have reviewed our approach to the derivation of mass formulas for stationary axisymmetric gravitating configurations with topological charges, based on the rod structure of solutions, and applied it to multi-black hole solutions in the Einstein-Maxwell theory, deriving the corresponding first law of black hole mechanics. This involves only the dynamical variables occurring in the first law for the Kerr-Newman solution, horizon area, angular momentum and electric charge, together with the conjugate surface gravity, horizon angular velocity and horizon electric potential, evaluated for all the Killing horizons, both black hole event horizons (timelike Killing horizons) and Misner strings (spacelike Killing horizons). To avoid ambiguities associated with the regularization of infinities, we have considered only asymptotically flat configurations with vanishing global NUT charge, so that Misner strings connect two adjacent black holes, but do not extend to infinity. The extension to the locally asymptotically flat case with a non-zero global NUT charge is straightforward. In this case, there will be an additional Misner string carrying the extra gravimagnetic flux to plus infinity, and another symmetrical Misner string carrying back the same flux from minus infinity, as discussed in \cite{smarrnut} in the case of the dyonic Kerr-Newman-NUT solution.

An important feature of our first law (\ref{gfirst}) is that it represents variation of mass (energy) in terms of variations of extensive (additive) quantities, namely angular momenta, component areas, and electric charges. Although we do not call this thermoynamics, which would be ambiguous for multi-black hole systems with different Hawking temperatures, this feature conforms to the general thermodynamic meaning of the energy functional. Most previous attempts (excepting that of \cite{krtous}) to formulate the first law for solutions with string singularities expressed energy variation in terms of a mixture of extensive and intensive quantities, which is thermodynamically incorrect. In particular, as we have shown here, attempts to preserve S-duality for multidyons lead to such incorrect variations.

A by-product of our analysis is the first general proof of the first law for NUTless rotating dyonic multi-black holes. In the presence of NUTs, adjacent black holes are connected by Misner strings carrying gravimagnetic flux, and also generically by Dirac strings carrying magnetic flux, as well as by negative tension cosmic strings which balance the forces between the black holes (except in the very special case of Majumdar-Papapetrou superpositions of extreme Reissner-Nordstr\"om black holes). Neither magnetic fluxes nor string tensions enter explicitly the first law, which involves only the dynamical variables of the Misner Killing horizons. When the NUT charges vanish, the Misner strings disappear, but the Dirac strings and cosmic strings remain. Their contributions to the first law, which were hidden behind the Misner string contributions in the NUTty case, become manifest in the NUTless case. It should be noted that the string tensions contribute to the first law, but not to the Smarr relations.

While the NUTless multi-dyon first law can be formally rearranged so that the magnetic flux contributions of the Dirac strings are replaced by black hole magnetic charge contributions, the resulting conjugate horizon magnetic potentials actually depend on the values of the fields on other horizons. This can be correlated with the fact, previously noted in \cite{Clement:2017otx}, that Dirac strings are massive, with a mass proportional to the magnetic flux through the string. The deep reason behind this breaking of electromagnetic S-symmetry is the difference in nature between electric and magnetic charges: the former are associated with conserved currents, the latter are topological. They cannot be unambiguously assigned to a black hole, but enter the general balance as independent entities associated with Dirac strings.

\section*{Acknowledgments}
The work of DG was partially supported by the Scientific and Educational School of Moscow State University “Fundamental and Applied Space Research”, and the Strategic Academic Leadership Program ``Priority 2030'' of the
Kazan Federal University.

\end{document}